\documentclass{emulateapj}

\shorttitle{Tracing shock precursors in L1448}
\shortauthors{I. Jim\'{e}nez--Serra et al.}

\begin{document}

\title{Tracing the shock precursors 
in the L1448--mm/IRS3 outflows}

\author{I. Jim\'{e}nez--Serra, J. Mart\'{\i}n--Pintado and A.
Rodr\'{\i}guez--Franco\altaffilmark{1}}
\affil{Instituto de Estructura de la Materia (CSIC), \\
Departamento de Astrof\'{\i}sica Molecular e Infrarroja, \\
C/ Serrano 121, E--28006 Madrid, Spain}

\and

\author{N. Marcelino}
\affil{Instituto de RadioAstronom\'{\i}a Milim\'etrica \\ 
Avda. Divina Pastora 7, N\'ucleo Central, 
E--18012 Granada, Spain}

\altaffiltext{1}{Escuela Universitaria de \'Optica,  
Departamento de Matem\'atica Aplicada (Biomatem\'atica),
Universidad Complutense de Madrid, 
Avda. Arcos de Jal\'on s/n, E--28037 Madrid, Spain}

\begin{abstract}

We present the detection of the SiO $\nu$=0 \textit{J}=2--1 
and \textit{J}=3--2 lines, and of the HCO 1$_{01}$--0$_{00}$
\textit{J}=3/2--1/2 \textit{F}=2--1  
line at ambient velocities towards the molecular 
outflows in L1448--mm and L1448--IRS3. This is the first detection of HCO in
a dark cloud. We have also
measured lines of H$^{13}$CO$^{+}$, H$^{13}$CN, HN$^{13}$C,
CH$_{3}$OH, and N$_2$H$^{+}$. While the HCO and the SiO lines have the
narrowest profiles with linewidths of $\sim$0.5$\,$km$\,$s$^{-1}$, the other lines
have widths of $\sim$1$\,$km$\,$s$^{-1}$. Towards L1448--mm, all lines except those
of SiO and HCO, show two distinct velocity components centered at 4.7
and  5.2$\,$km$\,$s$^{-1}$. HCO is only observed in the 4.7$\,$km$\,$s$^{-1}$ cloud,
and SiO in the 5.2$\,$km$\,$s$^{-1}$ component.
The SiO abundance is $\sim$10$^{-11}$ in the 5.2$\,$km$\,$s$^{-1}$ clouds,
one order of magnitude larger than in the 4.7$\,$km$\,$s$^{-1}$ component and in
other dark clouds. The HCO abundance is $\sim$10$^{-11}$, similar to that
predicted by the ion--molecule reactions models for the quiescent gas in
dark clouds. The large change in the SiO/HCO abundance ratio ($>$150)
from the 4.7 to the 5.2$\,$km$\,$s$^{-1}$ component, and the distribution and
kinematics of the SiO emission towards L1448--mm suggest that the ambient SiO
is associated with the molecular outflows. We propose that the narrow linewidths
and the abundances of SiO in the ambient gas are produced by the interaction
of the magnetic and/or radiative precursors of the shocks with the clumpy
pre--shocked ambient gas. 

\end{abstract}

\keywords{stars: formation --- ISM: individual (L1448) 
--- ISM: jets and outflows --- ISM: molecules
--- ISM: structure}

\section{Introduction}

It is well--known that the SiO emission traces the material which 
has been processed by shocks. The SiO profiles present very broad linewidths
that are centered at velocities different from that of the ambient cloud
(Mart\'{\i}n-Pintado, Bachiller, \& Fuente 1992). As silicon is heavily 
depleted onto dust grains in the cold quiescent gas, the SiO abundances are
extremely low in dark clouds ($\leq$10$^{-12}$; 
Ziurys, Friberg, \& Irvine 1989; Mart\'{\i}n-Pintado et al.~1992).
Nevertheless, Lefloch et al.~(1998) 
reported the first detection of bright and narrow (with the same linewidth of
the ambient gas) SiO lines at ambient velocities associated with the
molecular outflows in the dark cloud NGC~1333 (t$>$10$^{4}$ years). 
Lefloch et al.~(1998) suggested that the SiO emission arises from 
the postshock equilibrium 
gas after the interaction of powerful protostellar jets with local dense 
clumps. They concluded that the narrow SiO emission could
be only detected in particular objects characterized by its high degree of
clumpiness. Codella, Bachiller, \& Reipurth~(1999) 
confirmed the suggestion of Lefloch et al.~(1998) that the ambient SiO
emission arises from decelerated postshock material.
In this scenario one would not
expect to detect SiO ambient emission with narrow linewidths
in very young molecular outflows.

The L1448 outflow is a very young outflow with a dynamical age of only
$\sim$3500 years (Bachiller et al.~1990). Due its youth, this
object provides an excellent laboratory to study
the origin of the ambient SiO emission, because the time scale
is too short for the postshock gas to reach the equilibrium 
(Chi\`{e}ze, Pineau des For\^{e}ts, \& Flower~1998) and to decelerate
to the radial velocities of those of the ambient gas.

The formyl radical (HCO) has not been detected so far in
dark clouds (e.g. L183, where the derived HCO abundance is 
$<$2$\times$10$^{-10}$; Schenewerk et al.~1988). However, gas--phase chemistry
models predict even lower HCO abundances ($\sim$10$^{-11}$) than the previously
measured upper limits to this molecule (Leung, Herbst, \& Huebner~1984).
    
In this Letter we present the first detection of very narrow SiO 
and HCO emissions at ambient velocities in the 
L1448--mm/IRS3 molecular outflows. The narrow lines and the radial velocities
of the SiO emission do not seem to be related to decelerated postshock
equilibrium gas
but produced by the interaction of the magnetic and/or radiative
shock--precursors with the ambient pre--shocked clumpy medium. 
      
\section{Observations and Results}

The observations were carried out with the IRAM 30--m telescope 
at Pico Veleta (Spain). We observed the central position of L1448-mm
(the driving source of the highly collimated outflow)
and L1448--IRS3, and mapped the L1448--mm region observing towards the offsets
(20,0), (--20,0), (0,20), and (0,--20).  
Table~1 summarizes the parameters of the observed transitions.
The beam size of the telescope was $\sim$27'' at $\sim$90 GHz 
and $\sim$18''at $\sim$140 GHz. The 3~mm and the 2~mm
SIS receivers were tuned to single side band with image rejections $>$10~dB.
We observed simultaneously the SiO~$\textit{J}$=2--1,
H$^{13}$CO$^{+}$, HN$^{13}$C, and HCO species using the
frequency--switching mode with a frequency throw of 3~MHz. As spectrometers
we used VESPA, which provided a velocity resolution of
$\sim$0.14$\,$km$\,$s$^{-1}$. Typical system temperatures were 90--100~K.
We observed the rest of the species (H$^{13}$CN, CH$_{3}$OH, N$_2$H$^{+}$,
and the SiO~$\textit{J}$=3--2 transition) in position--switching mode.
The system temperatures and the spectral resolutions were typically of
120--350~K and 0.13--0.20$\,$km$\,$s$^{-1}$, respectively.
All the line intensities were calibrated in antenna temperature.

\begin{figure}
\epsscale{1.15}
\plotone{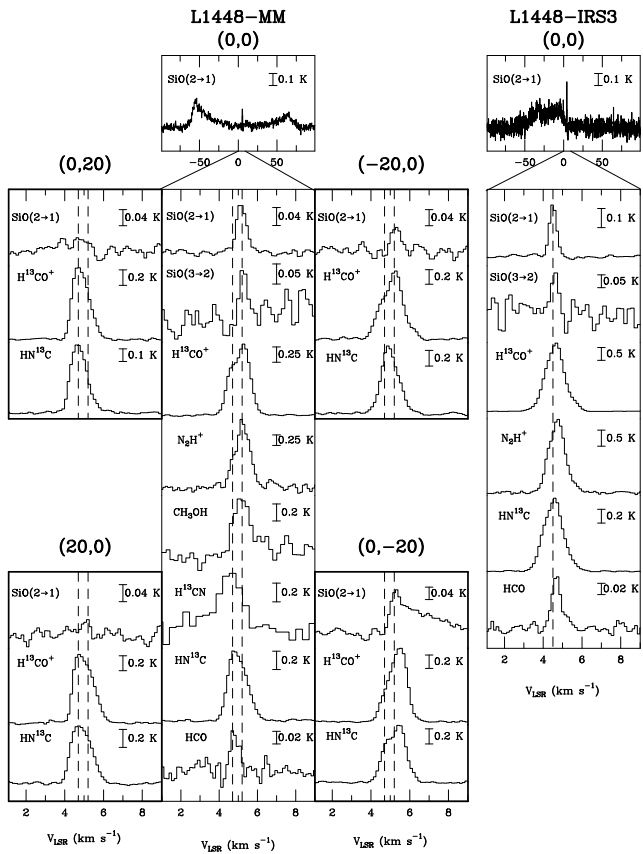}
\caption{Sample of line profiles for the different positions towards the
L1448--mm/IRS3 molecular outflows. The offsets, in arcseconds at the top of
the panels, refer to the position of L1448--mm
($\alpha$(2000)~=~03$^{h}$25$^{m}$38$_{.}^{s}$0,
$\delta$(2000)~=~30$^{\circ}$44$'$05$''$). The dashed vertical lines show the
radial velocities of the velocity components for which SiO and HCO have been
detected.}  
\label{fig1}
\end{figure}

Fig.~1 shows all the spectra measured for the L1448--mm map
and L1448--IRS3, and Table~1 shows the observed parameters for all
the lines measured towards the central positions in L1448--mm and L1448--IRS3.
For comparison, Fig.~1 shows, in the upper panels,
the profiles of the blueshifted and redshifted shocked gas with broad,
$\sim$50$\,$km$\,$s$^{-1}$, lines (see e.g., Mart\'{\i}n--Pintado et al.~1992).
The ambient profiles are shown in detail in the central panels of Fig.~1.
The SiO and HCO ambient emissions show very narrow linewidths of
$\sim$0.5$\,$km$\,$s$^{-1}$, smaller than those of the ambient cloud
molecules like H$^{13}$CO$^{+}$, HN$^{13}$C, and N$_2$H$^{+}$ which have
widths of $\sim$1$\,$km$\,$s$^{-1}$. There are appreciable differences between
the observed profiles of SiO, HCO, and the other molecules. While all ambient
molecules towards L1448--mm show double peaked profiles with velocity peaks
at 4.7 and 5.2$\,$km$\,$s$^{-1}$ (see vertical dashed lines in Fig.~1),
the HCO and SiO lines are single gaussian narrow profiles with peaks
at 4.7 and 5.2$\,$km$\,$s$^{-1}$ respectively (Table~1). VLA images of the
ambient NH$_3$ gas show that these two components correspond to distinct
molecular clumps with different spatial distribution associated with
the L1448--mm source (Curiel et al.~1999). 


\begin{figure}
\epsscale{1.15}
\plotone{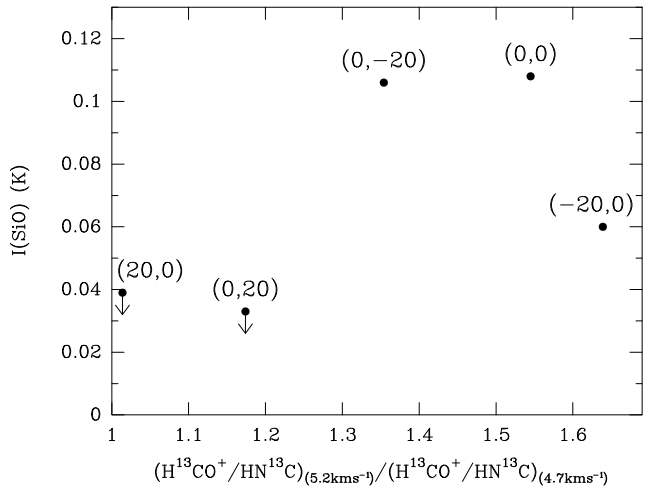}
\caption{Correlation between the narrow SiO emission intensity and the
ion (H$^{13}$CO$^{+}$) to neutral (HN$^{13}$C) intensity ratio
for the 5.2 and 4.7$\,$km$\,$s$^{-1}$ velocity components
detected towards the positions mapped in L1448--mm 
(the position offsets shown above the points are relative to L1448--mm).
The vertical arrows indicate upper limits to the SiO emission. 
Note the trend of the relative ion to neutral 
abundance to increase at the positions where SiO has been detected.
\label{fig2}}
\end{figure}

From our limited map, we notice that the detection of the SiO emission in the
5.2$\,$km$\,$s$^{-1}$ cloud is related to an increase of the ion abundance
with respect to neutrals. Fig.~2 illustrates this effect by representing
the SiO line intensities as a function of the ratio between the ion
(H$^{13}$CO$^{+}$) and the neutral (HN$^{13}$C) intensities for the 4.7 and
the 5.2$\,$km$\,$s$^{-1}$ components. Although the data are scarce,
the tendency of the SiO emission to increase with the ion to neutral
intensity ratio is clear. Since the lines are optically thin,
this shows that the larger SiO abundance in the ambient cloud is
associated with an enhancement of the ion abundance relative to the
neutrals. Furthermore, we find that the SiO profile towards L1448--mm (0,--20)
presents a substantial broadening (redshifted with respect to the ambient gas)
due to material which has already entered the shock. Although CH$_3$OH is also
a neutral molecule, its emission is shifted to the 5.2$\,$km$\,$s$^{-1}$
component. As discussed in section 4, these data are consistent with
the fact that the SiO and CH$_3$OH abundances are enhanced by grain mantle
destruction produced by shocks (Millar, Herbst, \& Charnley~1991).


\begin{deluxetable*}{ccccccc}
\tabletypesize{\scriptsize}
\tablecaption{Observed Parameters in $\textbf{L1448--mm/IRS3}$ outflows.
\label{tbl-1}}
\tablewidth{0pt}

\startdata
 & & & TABLE 1 & & & \\
     &  & $\textbf{L1448--mm}$ & & & $\textbf{L1448--IRS3}$ & \\ \hline\hline
Line & Intensity\tablenotemark{d} & $V_{LSR}$ & $\Delta v$ &
Intensity & $V_{LSR}$ & $\Delta v$ \\
      &  (K)      & (km s$^{-1}$) & (km s$^{-1}$) &  (K)      & (km s$^{-1}$) 
      & (km s$^{-1}$) \\ \hline
SiO($2\rightarrow1$)  &  0.108 (5) &   5.168 (9)   & 0.62 (2) &  0.275 (5)
&4.503 (4)   & 0.437 (9) \\
                      & $\leq$0.015 & $\sim$4.7  & & & & \\
SiO($3\rightarrow2$)  & 0.14 (4) &   5.29 (6) &  0.6 (2) & 0.173 (7) &
4.38 (5)   & 0.5 (1) \\
                      & $\leq$0.096 & $\sim$4.7  & & & & \\
H$^{13}$CO$^{+}$($1\rightarrow0$) & 1.150 (8) & 5.298 (7)   & 0.81 (1) &
2.58 (7) &   4.610 (1) &  1.159 (1) \\
                                  & 0.636 (8) & 4.585 (8) & 0.59 (2) & & & \\
N$_{2}$H$^{+}$ ($1\rightarrow0$)\tablenotemark{a} & 1.49 (6) & 5.25 (1) & 0.84
(3) & 2.22 (8) & 4.7 (1) & 1.0 (3) \\
                                     & 0.29 (6) & 4.61 (3) & 0.28 (7) & & & \\
HCO($1_{01}\rightarrow0_{00}$)\tablenotemark{b} & $\leq$0.03 & $\sim$5.2 &
& 0.094 (4) & 4.68 (5) & 0.5 (1) \\
                               & 0.059 (8) & 4.81 (4) & 0.49 (7) \\ 
HN$^{13}$C ($1\rightarrow0$)\tablenotemark{c} & 0.559 (5) & 5.16 (4) 
& 1.03 (3) & 1.02 (3) & 4.532 (2) & 1.187 (4) \\
                             & 0.428 (5) & 4.64 (1) & 0.62 (3) & & & \\
H$^{13}$CN ($1\rightarrow0$)\tablenotemark{c} & $\leq$0.30 & $\sim$5.2 &  \\
                                 & 0.40 (5) & 4.56 (9) & 1.5 (2) \\ 
CH$_{3}$OH($3_0\rightarrow2_0$ A$^{+}$) & 0.65 (6) & 5.20 (5) & 1.1 (1) \\
                                & $\leq$0.40 & $\sim$4.7 &  \\ \hline
\enddata

\tablenotetext{a}{Refers to the $\textit{F}$ = $0\rightarrow1$ hyperfine
component.}
\tablenotetext{b}{Refers to the $\textit{J}$ = $3/2\rightarrow1/2$,
$\textit{F}$ = $2\rightarrow1$ transition.}
\tablenotetext{c}{Refers to the $\textit{F}$ = $2\rightarrow1$ hyperfine
component.}
\tablenotetext{d}{Intensities are in antenna temperature. The limits correspond
to 3$\sigma$.}

\end{deluxetable*}

Towards L1448--IRS3, we have measured the strongest narrow
SiO emission in L1448. Again, the SiO and HCO lines
are narrower (with linewidths $\sim$0.5$\,$km$\,$s$^{-1}$) than those of the
ambient molecules. Toward this position all line
profiles are single peaked, but the SiO emission shows an
appreciable velocity shift relative to the emissions from ambient
molecules ($\sim$0.2$\,$km$\,$s$^{-1}$; see vertical line in Fig.~1). Like in
L1448--mm, this narrow emission is shifted in the same velocity direction as
the shocked material which suggests that the $''$ambient$''$ SiO
is related to shocks. 

\section{Abundances}

Assuming a kinetic temperature of 20~K (Curiel et al.~1999), Table~2 shows
the derived H$_2$ densities and the column densities of all molecules for
the central region of L1448--mm and L1448--IRS3. The H$_2$ densities derived
from the SiO~$\textit{J}$=2--1 and $\textit{J}$=3--2 lines are a few
10$^{5}$--10$^{6}$~cm$^{-3}$. If we consider a HCO$^{+}$ fractional abundance
of $\sim$10$^{-8}$ (Irvine et al.~1987), the derived SiO abundance
towards L1448--mm (0,0), (--20,0), and (0,--20)
is $\sim$10$^{-11}$ for the 5.2$\,$km$\,$s$^{-1}$ component, an order of
magnitude lower than what Lefloch et al.~(1998) found in NGC~1333.
The estimated upper limit to the SiO abundance for the 4.7$\,$km$\,$s$^{-1}$
cloud is $\leq$10$^{-12}$ at L1448--mm (0,0), of the same order as that
obtained in the cold quiescent clouds TMC1 and B1 (Ziurys et al.~1989;
Mart\'{\i}n--Pintado et al.~1992). Therefore, the SiO abundance in the
5.2$\,$km$\,$s$^{-1}$ component has been enhanced by more than one order
of magnitude with respect to the quiescent gas of the 4.7$\,$km$\,$s$^{-1}$ cloud. 

On the other hand, HCO has not been detected for the 5.2$\,$km$\,$s$^{-1}$
component (Fig.~1), with an upper limit of
$\chi$(HCO)$\leq$4$\times$10$^{-12}$ (Table~2).
For the 4.7$\,$km$\,$s$^{-1}$ cloud, we derive a HCO abundance of few 10$^{-11}$.


\begin{deluxetable*}{lccccccccc}
\tabletypesize{\scriptsize}
\tablecaption{Derived parameters in $\textbf{L1448--mm/IRS3}$ outflows.
\label{tbl-2}} 
\tablewidth{0pt}
\startdata

 & & & & TABLE 2 & & & & &  \\
\\ \hline\hline 

Source & H$_2$ density & &  &Column & density & (10$^{12}$cm$^{-2}$)&  
& $\chi$(SiO) & $\chi$(HCO) \\
\cline{3-8}
& 10$^{5}$ (cm$^{-3}$) & SiO & HCO & H$^{13}$CO$^{+}$ & HN$^{13}$C &
CH$_3$OH & N$_2$H$^{+}$ & & \\ \hline
L1448-mm (5.2 km s$^{-1}$) & $\sim$10 & 0.2 & $\leq$0.03 & 1.0 & 1.0 & 80 & 20  &
1$\times$10$^{-11}$ & $\leq$4$\times$10$^{-12}$ \\
L1448-mm (4.7 km s$^{-1}$) & $\sim$10 & $\leq$0.004 & 0.2 & 0.4 & 0.5 & 
$\leq$30 & 1.1  & $\leq$1$\times$10$^{-12}$ & 6$\times$10$^{-11}$ \\
L1448-IRS3 & 2.4 & 0.3 & 0.4 & 2.8 & 1.7 & & 40 & 1$\times$10$^{-11}$ & 
2$\times$10$^{-11}$\\  \hline

\enddata

\tablecomments{Abundance upper limits were estimated from
the 3$\sigma$ level. These parameters have been derived for the central
region in each source.}
\end{deluxetable*}

\section{Discussion}

HCO is predicted to be produced in dark clouds through ion--molecule
reactions with abundances of 10$^{-11}$ (Leung et al.~1984). Our detection
of HCO and the abundances derived for this molecule ($\sim$10$^{-11}$;
Table~2) are in agreement with these models, suggesting that the
4.7$\,$km$\,$s$^{-1}$ material traces the quiescent gas. The SiO/HCO abundance
ratio changes by more than two orders of magnitude from the 4.7 to the
5.2$\,$km$\,$s$^{-1}$ component of the ambient gas, indicating that the formation
of both molecules are related to different mechanisms.

The spatial distribution and the kinematics of the narrow SiO emission suggest
that this emission is associated with the molecular outflows towards L1448--mm
and L1448--IRS3. However, the lines are very narrow and with the radial
velocities of the ambient gas. We can rule out the idea that the ambient
SiO emission is generated in the postshock equilibrium gas since
the time scales for the shocked gas to decelerate to the
ambient velocities are much larger ($\geq$10$^{4}$ years) 
than the dynamical age, and the SiO has not been detected in the ambient
4.7$\,$km$\,$s$^{-1}$ cloud.

In the following, we explore the possibility that the SiO ambient emission
is a signature of the shock precursors associated with the jets driven by the
young stars. There is a tendency for the ambient SiO emission to arise only
from regions where the relative abundance of the ions to neutrals 
is larger than in the quiescent clumps (see Fig.~1).
This indicates that the ions could have slipped to the 5.2$\,$km$\,$s$^{-1}$
component from the neutrals of the 4.7$\,$km$\,$s$^{-1}$ cloud,
as expected if this narrow emission were produced by the magnetic precursor
associated with C--type shocks. The magnetic precursor
would force the plasma to stream through the neutral gas
ahead the jump front, generating a velocity decoupling between the charged
and neutral fluids (Draine~1980). As the charged particles were compressed
in the first stages of the magnetic precursor, the ion
density would initially increase in the 5.2$\,$km$\,$s$^{-1}$
component with respect to the quiescent 4.7$\,$km$\,$s$^{-1}$ cloud
(see e.g. Flower et al.~1996).

Pilipp, Hartquist, \& Havnes~(1990) and Pilipp \& Hartquist~(1994)
have considered grains as a separated fluid from the ion, neutral, and
electron components. The behavior of the grains changes depending on
several parameters like the $H_2$ density, mean charge, and grain size
distribution. There are cases where the charged grains are coupled to
the ions, leading to a small grain--neutral velocity drift just 
before the shock starts. The Chi\`{e}ze et al.~(1998)
time--dependent C--shock model shows that, even for time scales
of $<$10$^{4}$$\,$years, this velocity drift generates the sufficient
friction to produce
the ejection of small amounts of silicon from grains into the gas phase in the
5.2$\,$km$\,$s$^{-1}$ cloud through sputtering
(Flower et al.~1996; Caselli, Hartquist, \& Havnes~1997).
Since these time scales are rather short, we should expect that silicon is
released directly in the form of SiO, producing an abundance of 10$^{-10}$
or 10$^{-11}$ (Flower et al.~1996; Caselli et al.~1997).
The lack of a velocity
shift between the SiO and the ions for the 5.2$\,$km$\,$s$^{-1}$
component also indicates that we are observing the very early phases of the
interaction of the shock precursor. This is supported by the data 
at the position (0,-20) where the broader SiO profile (see Fig.~1)
suggests that in this case a large fraction of the neutrals has already
entered the shock.

On the other hand, the detection of [Si II] 34.8$\,$$\mu$m emission in
L1448--IRS3 (Nisini et al.~2000) shows the existence of a photon--dominated
region (PDR) which could be related to the radiative precursor of the J--type
shocks (Shull \& McKee~1979; Hollenbach \& McKee~1989).
In fact, the C--type shock models of Chi\`{e}ze et al.~(1998) predict a jump
discontinuity in the neutral flow in the early stages for time scales 
$\leq$10$^{4}$~years. The UV radiation field of the
radiative precursor associated with such J--shock could 
affect the chemistry of the pre--shocked medium.
It might be possible that the ambient SiO emission would be produced
by the radiative precursor through the photodesorption of Si from the grains
(Schilke et al.~2001). However, the SiO/HCO abundance ratio in the
5.2$\,$km$\,$s$^{-1}$ velocity component is very different from what is observed
in PDRs (Schilke et al.~2001).

In summary, the linewidths, velocities, abundances, and spatial distribution
of the SiO in L1448--mm/IRS3 could be explained by the radiative and/or
magnetic shock precursors associated with the jets driven by protostars. Based
on all these results, we speculate that we have probably detected the
first evidence of the chemical effects on the ambient cloud produced by the
precursors of interstellar shocks. Higher angular resolution observations
of SiO, HCO, and tracers of the ambient gas are needed to clearly establish
the origin of the ambient emission of SiO in L1448.

\acknowledgments

This work has been supported by the Spanish MCyT under projects
number ESP2002--01627, AYA2002--10113--E and AYA2003--02785--E.   
We thank J.R. Goicoechea for useful comments and suggestions. We also
thank an anonymous referee for helping us to improve the manuscript.


\begin{thebibliography}{}
\bibitem[Bachiller et al. 1990]{bac90} Bachiller, R., 
Cernicharo, J., Mart\'{\i}n--Pintado, J., Tafalla, M., 
\& Lazareff, B. 1990, \aap, 231, 174
\bibitem[Caselli et al. 1997]{cas97} Caselli, P., Hartquist, T. W., \&
Havnes, O. 1997, \aap, 322, 296
\bibitem[Chi\`{e}ze et al. 1998]{chi98} Chi\`{e}ze, J.--P., Pineau des
For\^{e}ts, G., \& Flower, D. R. 1998, MNRAS, 295, 672 
\bibitem[Codella et al. 1999]{cod78} Codella, C.,
    Bachiller, R., \&  Reipurth, B. 1999, \aap, 343, 585
\bibitem[Curiel et al. 1999]{cur99} Curiel, S., 
Torrelles, J.M., Rodr\'{\i}guez, L. F., G\'omez, J. F., 
\& Anglada, G. 1999, \apj, 527, 310
\bibitem[Draine 1980]{dra80} Draine, B. T. 1980, \apj, 241, 1021
\bibitem[Flower et al. 1996]{flo96} Flower, D. R., Pineau des
For\^{e}ts, G., Field, D., \& May, P. W. 1996, MNRAS, 280, 447 
\bibitem[Hollenbach \& McKee 1989]{hol89} Hollenbach, D., \& McKee, C. F.
1989, \apj, 342, 306
\bibitem[Irvine et al. (1987)]{irv87} Irvine, W. M.,
    Goldsmith, P. F., \& Hjalmarson, A.  1987, Interstellar processes.
    Hollenbach, D. J., Tromson, H. A. (eds.) Reidel Dordrecht, p. 561
\bibitem[Lefloch et al. (1998)]{lef98} Lefloch, B.,
     Castets, A., Cernicharo, J., \& Loinard, L. 1998, \apjl, 504, L109
\bibitem[Leung et al. (1984)]{leu84} Leung, C. M., Herbst, E., \& Huebner,
W. F. 1984, ApJS, 56, 231
\bibitem[Mart\'{\i}n--Pintado et al. (1992)]{mar92}
Mart\'{\i}n--Pintado, J., Bachiller, R., \& Fuente, A. 1992, \aap, 254, 315
\bibitem[Millar et al. 1991]{mil91} Millar, T. J., Herbst, E., \& Charnley,
S. B. 1991, \apj, 369, 147
\bibitem[Nisini et al. (2000)]{nis00}
Nisini, B., Benedettini, M., Giannini, T., Codella, C., Lorenzetti, D., Di
Giorgio, A. M., \& Richer, J. S. 2000, \aap, 360, 297
\bibitem[Pilipp et al. (1990)]{pil90} Pilipp, W., Hartquist, T. W.,
    \& Havnes, O.  1990, MNRAS, 243, 685
\bibitem[Pilipp \& Hartquist 1994]{pil94} Pilipp, W., \& Hartquist, T. W.
1994, MNRAS, 267, 801
\bibitem[Schenewerk et al. (1988)]{sch88} Schenewerk, M. S., Snyder, L. E.,
Hollis, J. M., Jewell, P. R., \& Ziurys, L. M. 1988, \apj, 328, 785
\bibitem[Schilke et al. (2001)]{sci01} Schilke, P., Pineau des For\^{e}ts, G.,
Walmsley, C.M., \& Mart\'{\i}n--Pintado, J. 2001, \aap, 372, 291
\bibitem[Shull \& McKee (1979)]{shu79} Shull, J. M., \& McKee, C. F. 1979,
\apj, 227, 131
\bibitem[Ziurys et al. (1989)]{Ziu85} Ziurys, L. M., Friberg, P., \& Irvine,
W. M. 1989, \apj, 343, 301
\end{thebibliography}
\end{document}